\documentstyle[12pt]{article}

\begin{document}
\begin{center}
{\Large \bf GEODESIC CURVES ON QUANTIZED MANIFOLDS}\\
\vspace{1cm}
{\bf V. Milani$^{1,2,*}$ , A.Shafei Deh Abad$^{1,3,**}$} \\
\vspace{3mm}
{\it $^1$Institute for Studies in Theoretical Physics and 
Mathematics, P.O.Box 5531,Tehran, Iran.}\\
{\it $^2$Department of Mathematics, University of Shahid 
Beheshti, Tehran, Iran.}\\
{\it $^3$Department of Mathematics, University of Tehran, Tehran, Iran.}\\
{\it $^*$e-mail: milani@physics.ipm.ac.ir}\\
{\it $^**$e-mail: shafei@physics.ipm.ac.ir}\\

\vspace{7mm}

\end{center}
\vspace{3mm}
\begin{abstract}
A general definition of the curves and geodesics associated with a given 
connection on a quantized manifold is given. In the particular case 
of the functional quantization we define geodesics in the same way 
as in the classical case and we will show that the two definitions are
compatibles. As an example we examine our results for the quantum Manin plane.
\end{abstract}
\newpage
{\bf 0.Introduction}

Recently there have been so many works on noncommutative geometry and 
the related topics by physicists and mathematicians (A. connes 1986, 1995,  
Dubois-Violette 1994, Mourad 1994, Wess \& Zumino 1990). In this paper 
following the above works we give the definition of geodesics on quantized 
manifolds and find them for the particular case of Manin plane.

{\bf 1.Linear Connections}

We apply the general definition of a linear connection in noncommutative 
geometry proposed by A. Connes [1] in order to generalize the notion of 
geodesic curves on quantized manifolds.

First we recall the definition of a linear connection in the commutative 
case. If $M$ is a smooth manifold and $C^{\infty}(M)$ the algebra of 
smooth functions on $M$, then a linear connection on a smooth complex 
vector bundle $E$ on $M$ is given by the C-linear map $$\nabla : \Gamma(E)
\rightarrow \Omega^1(M) \otimes_{C^{\infty}(M)} \Gamma(E)$$
where $\Gamma(E)$ is the finitely generated projective $C^{\infty}(M)$-module of sections
of $E$ and $\Omega^1(M)$ is the $C^{\infty}(M)$-module of first order differential
forms on $M$. The linear map $\nabla$ satisfies the following Leibniz rule
\begin{equation}
\nabla(f\xi)=df \otimes \xi + f \nabla(\xi)
\end{equation}
for each $f \in C^{\infty}(M)$ and $\xi \in \Gamma(E).$

The fact that $C^{\infty}(M)$ is a commutative algebra allows us to use the
relation $\nabla(f\xi)=\nabla(\xi f)$ to write (1) as
\begin{equation}
\nabla(\xi f)=\sigma (\xi \otimes df) + (\nabla \xi)f
\end{equation}
where $\sigma$ is the permutation operator.

Thanks to the well known equivalence $E \leftrightarrow \Gamma(E)$, between
the category of smooth vector bundles on $M$ and the category of finitely 
generated projective $C^{\infty}(M)$-modules, each linear connection 
can be considered on finitely generated projective modules. The latter 
is suitable enough to be generalized to the noncommutative case.

In noncommutative geometry the role of $C^{\infty}(M)$ is replaced by a
noncommutative associative algebra ${\cal A}$ and $\Gamma(E)$ is replaced by
a finitely generated projective ${\cal A}$-module $P$. 
A linear connection in general case can now be defined on $P.$ For our purpose 
we take
$P = \Omega^1({\cal A})$, the ${\cal A}$-module of first order differential 
forms on ${\cal A}$ and a linear connection on $\Omega({\cal A})$ is a linear
map $$\nabla : \Omega({\cal A}) \rightarrow \Omega({\cal A}) \otimes_{{\cal A}}
\Omega({\cal A})$$
satisfying (1) for $f \in {\cal A}$ and $\xi \in \Omega({\cal A}).$

In noncommutative case in general $\nabla(f \xi) \neq \nabla(\xi f)$ and so
we impose relation (2) for a suitable substitution for $\sigma.$
From now on $I \subset R$ is an interval.

{\bf 2.Geodesic Curves in Commutative Geometry}

To each linear connection in commutative geometry, there correspods geodesic 
curves. In what follows we try to give an equivalent definition for this 
concept appropriate for the quantized contexts.

On a smooth manifold $M$ with a fixed linear connection $\nabla_{M}$ and 
associated christofel symbols $\Gamma_{ij}^k$, the geodesic curves are 
defined locally as the solutions of the following differential eqaution
\begin{equation}
\frac{d^2{x^k}}{dt^2} + \Gamma_{ij}^k \frac{dx^i}{dt} \frac{dx^j}{dt} = 0.
\end{equation}
Direct calculation shows that the condition for a parametrized curve $\gamma$
on $(M,\nabla_{M})$ to be a geodesic curve, is eqivalent to the relation
\begin{equation}
(\gamma^* \otimes \gamma^*) \circ \nabla_{M} = \nabla_{R} \circ \gamma^*
\end{equation}
where $\nabla_{R}$ is the extension of the standard linear connection 
$\delta(f)=df$ given by $\nabla_{R}(f\xi) = df \otimes \xi$ for 
$f \in C^{\infty}(I)$ and $\xi \in \Omega^1(I).$

This relation can be used to generalize geodesics in quantized case.

{\bf 3.Curves and Geodesics on Quantized Manifolds} 

Let ${\cal A}$ be a quantization of the commutative algebra $B.$ More 
precisely ${\cal A}$ is a noncommutative algebra over a subring of the formal 
power series
in $\lambda$, having the character $\psi$ and $\phi:{\cal A} \rightarrow B$ 
is a $\psi$-homomorphism
i.e. $\varphi(\lambda f)= \psi(\lambda) \varphi(f).$ 
More details can be found 
in [2,3].
Now any $\psi$-homomorphism $$\gamma^* : {\cal A} \rightarrow C^{\infty}(I)$$
is called a curve on the quantized algebra ${\cal A}$. 

We say that a curve $\gamma: {\cal A} \rightarrow C^{\infty}(I)$ is maximal
if for each interval $J \subset R$ with $I \subset J$ and each curve 
$\gamma': {\cal A} \rightarrow C^{\infty}(J),$ the relation $i^* \circ 
\gamma' = \gamma$ implies $\gamma = \gamma',$ where $i$ is the inclution map.

In the following $x^i$ are linearly independent generators of the algebra
${\cal A}.$ 
Given a connection $\nabla$ on $\Omega^1({\cal A})$ the geodesic curves are 
defined to be those curves $\gamma^*$ on ${\cal A}$ satisfying the relation (4)
with the same symbol $\gamma^*$ as the extension of $\gamma^*$ to $\Omega({\cal A})$ by 
$\gamma^*(df) = d \gamma^*(f)$ for $f \in {\cal A}.$

As in the commutative case we find that the relation (4) is equivalent to 
the equation
\begin{equation}
\frac{d^2 \gamma^*(x^k)}{dt^2} + \gamma^*(\Gamma_{ij}^k)\frac{d \gamma^*(x^i)}{dt}
\frac{d \gamma^*(x^j)}{dt} = 0
\end{equation}
where $\Gamma_{ij}^k \in {\cal A}$ are the associated christofel symbols for 
$\nabla.$ We emphasize that by knowing the christofel symbols 
$\gamma^*(\Gamma_{ij}^k)$ is determind in terms of $\gamma^*(x^i), \gamma^*(x^j).$
Therefore this equation is the usual second order differential equation 
in terms of $\gamma^*(x^i).$
Since according to the functional quantization [3], the quantized manifold $M$
is the set 
of its points, so curves on these manifolds have parametrized representations
as in the classical case. In the case of smooth functional quantization, to 
each such parametrized curve $\gamma : I \rightarrow M$ there 
corresponds a $\psi$-homomorphism $$\gamma^* : {\cal A} \rightarrow 
C^{\infty}(I)$$ given by $\gamma^*(f) = f \circ \gamma|_{\lambda=\lambda_0}.$ 
In this case we can define the geodesic curves locally as the solutions of 
the equation 
\begin{equation}
\frac{d^2x^k}{dt^2} + \Gamma_{ij}^k \frac{dx^i}{dt} 
\frac{dx^j}{dt} = 0
\end{equation}
with the associated christofel symbols for $\nabla$ on $\Omega({\cal A})$ as 
functions of $t.$
It is seen that the solutions of this equation $\gamma(t)=(x^i(t))_i$  are 
related to the solutions of the equation (5) by 
$\gamma^*(f)=f \circ \gamma |\lambda=\lambda_0.$ So the two definitions are compatible.

{\bf 4.The Quantum Plane}

In this section we use our definitions of curves and geodesics for the case
of quantum plane.
The algebra of differential forms on the quantum plane $\Omega = \Omega^0 
\oplus \Omega^1 \oplus \Omega^2$ has generators $x,y,dx,dy$ satisfying the 
commutation relations
$$xy = q yx, x dx = q^2 dx x, y dy = q^2 dy y, dy dx + q dx dy = 0,$$
$$ x dy = q dy x + (q^2-1)dx y, dx dx = dy dy = 0.$$
A linear connection $\nabla$ on this plane compatible with the above relations 
is given by [4] as $$\nabla(dx^i) = \mu^4 x^i \theta \otimes \theta$$
where $x^1=x,x^2=y$ and $\theta = x dy - q dy x.$ The parameter $\mu$ is
such that in the classical limit $q \rightarrow 1$ tends to zero.

The associated christofel symbols are found to be 
$$\Gamma_{11}^1=- \mu^4xy^2, \Gamma_{12}^1=\mu^4q^{-1}x^2y, 
\Gamma_{21}^1=\mu^4x^2y,$$
$$\Gamma_{22}^1=- \mu^4q^{-1}x^3, \Gamma_{11}^2=- \mu^4qy^3, \Gamma_{12}^2=
\mu^4q^{-2}xy^2,$$$$ \Gamma_{21}^2=\mu^4q^{-1}xy^2, 
\Gamma_{22}^2=\mu^4q^{-3}x^2y.$$
Since in this case $\gamma^*(\Gamma_{ij}^k) = 0$ the equation (5) becomes 
$$\frac{d^2 \gamma^*(x^i)}{dt^2} = 0$$
which has the solutions 
\begin{equation}
\gamma^*(x^i)(t) = a^it + b^i 
\end{equation}
where $a^i, b^i$ are numbers.

Since we can consider the Manin plane as the functional quantization of R$^2$,
by using the above christofel symbols we see that the solutions of (4) are  
of the following form
$x^i(t)={a^i}_0(q)+{a^i}_1(q)t+ \mu^4 \varphi (t),$
where $\varphi \in C^{\infty}(R).$
Clearly these solutions satisfy (7).

\end{document}